\documentclass[doublespacing]{elsart}
\newcommand{\DDDA}{\textrm{$D$}\&{} }
\newcommand{\DDD}{\textrm{$D$}{} }
\usepackage{graphicx}
\usepackage{amsbsy}
\usepackage{psfrag}
\usepackage{amssymb}
\begin{document}
\begin{frontmatter}

\title{
Limits, discovery and cut optimization for a Poisson process with 
uncertainty in background and signal efficiency: TRolke 2.0
}

\author[label1]{J.~Lundberg}\author[corr]{},\author[label2]{J.~Conrad},\author[label3]{ W.~Rolke},\author[label3]{A.~Lopez }
\address[label1]{CERN, CH-1211 Gen\`eve 23, Switzerland}
\address[label2]{Stockholm University,
Oskar Klein Centre for Cosmoparticle Physics
AlbaNova University Centre,
SE-10961 Stockholm, Sweden}
\address[label3]{
University of Puerto Rico-Mayaguez, Car 2,
Mayaguez, Puerto Rico 00681 (USA)
Mayaguez PR 00681
}
\corauth[corr]{Corresponding author: johan.lundberg@cern.ch} 

\begin{abstract}
A C++ class was written for the calculation of frequentist confidence
intervals using the profile likelihood method. Seven combinations of
Binomial, Gaussian, Poissonian and Binomial uncertainties are
implemented. The package provides routines for the calculation of upper and
lower limits, sensitivity and related properties. It also supports
hypothesis tests which take uncertainties into account. It can be used in
compiled C++ code, in Python or interactively via the ROOT 
analysis framework.
\end{abstract}

\end{frontmatter}

\label{}
\begin{itemize}
\item{\it Title of Program:} TRolke version 2.0
\item{\it Program available from: } CPC Program Library, ...
\item{\it Licensing provisions:} MIT License
\item{\it Computer for which the program is designed:} Unix, GNU/Linux, Mac
\item{\it Operating Systems under which the program has been tested:} Linux
  2.6 (Scientific Linux 4 and 5, Ubuntu 8.10) , Darwin 9.0 (Mac-OS X 10.5.8)
\item{\it Programming Language used:} ISO C++
\item{\it Memory required to execute with typical data:} $\sim$ 20 MB, 
\item{\it No. of bytes in distributed program, including initialization file, etc..} 1 MB
\item{\it Distribution Format:} tar file
\item{\it Keywords:} confidence interval calculation, systematic
  uncertainties, profile likelihood
\item{\it Nature of the Physical Problem: } The problem is to calculate a
  frequentist confidence interval on the parameter of a Poisson process with
  statistical or systematic uncertainties in signal efficiency or background.
\item{\it Method of solution:} Profile likelihood method, Analytical
\item{\it Typical Running Time:} $<10^{-4}$ seconds per extracted limit.
\end{itemize}

\newpage

\tableofcontents

\newpage

\section{Introduction and scope}

Routines were written for the calculation of frequentist confidence intervals
using the profile likelihood method. The package provides routines for the
calculation of upper and lower limits, average limits (sensitivity) and
related properties, taking uncertainties in background estimate and signal
efficiency into account.  The implementation considers seven different
statistical models with different combinations of Binomial, Gaussian,
Poissonian or no uncertainties. For example in the Gaussian background case,
our package derives upper and lower limits on the signal strength for a
Poisson process with Gaussian background expectation ${b}\pm\delta{}{b}$. It is
also possible to construct hypothesis tests which take uncertainties into
account. The statistical problems are treated using the $\emph{Profile Likelihood
  method}$.

The package provides a $\textsc{C++}$ class with accompanying examples. It can
be used in compiled code, interactively via the ROOT \cite{ROOT} analysis
framework, and from Python. This is TRolke version 2.0. It adds to version 1
(implemented in Fortran and in C++): hypothesis tests, a reworked
user interface, documentation, examples and python support.

This paper is organized as follows. First, the profile likelihood methods is
summarized, section \ref{pfmethod} ; second, it is shown how our routines can
be used for optimization of statistical discovery or limit setting power,
section \ref{optimization}. The means for specification of the statistical
model, and in general the class interface are described in section
\ref{classdoc}.

\section{The profile likelihood method}
\label{pfmethod}

Frequentist limits are constructed from data such that when repeated with new
data the limits \emph{cover} the fixed but unknown parameter value $\pi$ with
a frequency which converges to the requested probability, the confidence level
$1-\alpha$. Limit calculation methods are often based on the inversion of an
hypothesis test, as described in e.g.
\cite{Feldman:1997qc}\cite{Kendall:91}\cite{Neyman:1937a}, and we follow the
same scheme. Classical hypothesis tests investigate the validity of a default
hypothesis, the \emph{null hypothesis} $\mathcal{H}_\textrm{0}$; that an
examined sample of data is compatible with background and we call the
complementary hypothesis $\mathcal{H}_\textrm{1}$ a \emph{discovery}.  The
profile likelihood method is based on the likelihood ratio tests statistic now
described. For some observable $X$, let us assume a probability
density function $f(X_i|\boldsymbol{\pi},\boldsymbol{{b}})$ depending on
$k$ parameters $\boldsymbol{\pi}= \{\pi_1,\ldots,\pi_k\}$ of interest to the
researcher (such as the strengths of different signal sources), and $l$
additional nuisance parameters $\boldsymbol{{b}} =
\{{b}_1,\ldots,{b}_l\}$ (such as the strength of different background
sources). For a set of $n$ independent observations $\boldsymbol{X} =
\{X_1,\ldots,X_n\}$ the likelihood is
$$
L(\boldsymbol{\pi},\boldsymbol{{b}}|\boldsymbol{X}) = \prod_{i=1}^{n} f(X_i|\boldsymbol{\pi},\boldsymbol{{b}}).
$$

The likelihood ratio test statistic is defined as 
$$
\lambda(\boldsymbol{\pi_0} | \boldsymbol{X}) =
\frac{
\textrm{sup}\{L(\boldsymbol{\pi},\boldsymbol{{b}}|\boldsymbol{X}) ; 
\boldsymbol{\pi} = \boldsymbol{\pi}_0, \boldsymbol{{b}} \}}{
\textrm{sup}\{L(\boldsymbol{\pi},\boldsymbol{{b}}|\boldsymbol{X}) ; 
\boldsymbol{\pi}, \boldsymbol{{b}} \}},
$$
where the denominator is the likelihood maximized over the whole
$\{\boldsymbol{\pi}, \boldsymbol{{b}}\}$ space, while the nominator is
maximized over the more restrictive null hypothesis space $\{\boldsymbol{\pi}
= \boldsymbol{\pi}_0, \boldsymbol{b}\}$. The likelihood ratio $\lambda$
is also known as the \emph{profile likelihood} and is a stochastic function
explicitly depending on the data (and the null hypothesis) but not the
nuisance parameters. 

In general the inversion of a test to find the confidence region requires
scanning over all possible signals, as described for example in
\cite{Feldman:1997qc}. Our routines instead make use of a very powerful result
from mathematical statistics, that 
under some general conditions the distribution of $-2 \log \lambda$ converges
to a chi-square distribution with $k$ degrees of freedom.  Although these
conditions are not satisfied in the problem considered here it has been shown
that its performance is surprisingly good, especially when, as here, nuisance
parameters are included. The statistical performance of the Profile likelihood
method is studied in Ref.\cite{trolkenull:2004}. 

\section{Analysis optimization for optimal limits or discovery power}

In this section we describe how our routines are used for optimization of
analysis cuts, with the figure of merit being either stringent limits (in
case the signal is expected to be weak), or probability for discovery (if the
signal is expected to be strong).
\label{optimization}

\subsection{Analysis optimization for stringent limits in case of vanshing signal}

\label{mrf}
When a signal is expected to be weak enough so that significant
discovery is unlikely, it is relevant to optimize the analysis for
optimal limit setting power.
This can be done by assuming no signal and
minimizing the so-called \emph{sensitivity}. For example with a 90\% confidence level
(that is, $\alpha\!\!=$10\%), let us denote a calculated upper limit
$s_{90}$. The sensitivity of the experiment is defined as the average
upper limit in case of vanishing signal;
\begin{equation}
  \overline{s}_{90}(b) = \sum_{x=0}^\infty P(x,b)s_{90}(x,b),
\end{equation}
where $P(x,b)$ is the Poisson probability of observing $x$ events for
background expectation $b$, in absence of signal. 
For finding the optimal analysis cut we can assume without loss of generality
that the background and signal expectations are monotonically decreasing
functions of a cut $c$: $s(c)=\mu_s \epsilon_s(c)$, and $b(c)$. The constant
$\mu_s$ is the assumed normalisation of the signal at some arbitrary ``no
cut'' level so that all uncertainties in the signal rate expectation are
attributed to the signal detection efficiency $\epsilon_s$.

As an example, let's consider an energy dependent spectrum of particles
probed by a particle detector. For the physical \emph{test spectrum} 
\begin{equation}
  \frac{\mathrm{d}\Phi_\textrm{test}(E)}{dE} \equiv a_{\textrm{test}} \frac{\mathrm{d}\phi(E)}{dE},
\end{equation}
the expected number of observed signal events is 
\begin{equation}
s_\textrm{test} =  a_{\textrm{test}} \cdot T \int \mathrm{d}\Omega \int \frac{ \mathrm{d}\phi(E)}{dE} \sigma(E) \textrm{d}E. 
\end{equation}
The cross section $\sigma$ determines the detection efficiency which is
now a function of the energy $E$, and $T$ is the exposure time. 
For the observation of $x$ events the \emph{model rejection factor} $\xi(x)$
is defined as
\begin{equation}
  \xi(x) = s_{90}(x_{{0}},b)/s_{\textrm{test}}.
\end{equation}
%where $s_{\textrm{test}}$ is the
%expectation number of signal events for an assumed test signal. 
The
upper limit can be written in terms of the test signal
\begin{equation}
 \frac{\mathrm{d}{\Phi}_{90}(E)}{dE}  =  {\xi}(x) \cdot  \frac{\mathrm{d}\Phi_\textrm{test}(E)}{dE}
\label{limiteqn},
\end{equation}
and the average limit on the signal strength, set by repeated independent
experiments in case of vanishing signal is
\begin{equation}
\label{mrfanalogy}
 \frac{\mathrm{d}\overline{\Phi}_{90}(E)}{dE}  =  \overline{\xi} \cdot  \frac{\mathrm{d}\Phi_\textrm{test}(E)}{dE},
\end{equation}
where $ \overline{\xi} = \overline{s}_{90}(b)/s_\textrm{test}$ is called the
\emph{model rejection potential}. 

Our package provides $\overline{s}_{90}$ through the method
\textsf{GetSensitivity(\DDDA{} $s_\textrm{L}$, \DDDA{} $s_\textrm{U}$)}, and the upper limit
$s_{90}$ (and the lower limits) through \textsf{GetLimits(\DDDA{} $s_\textrm{L}$,
  \DDDA{} $s_\textrm{U}$)}, where $D$ indicates a double precision value.

\subsection{Hypothesis testing with uncertainties}

\label{hyptest}
In order to reject
$\mathcal{H}_\textrm{0}$ with significance $\alpha$, the number of observed
events $x_0$ must be equal to or higher than a \emph{critical number}
$x_{\textrm{c}}(b)$, where $b$ is the background expectation. The significance
$\alpha$ is the probability of observing $x_{\textrm{c}}$ or more events from
a stochastic background with mean $b$ assuming vanishing signal. 

The part of sample space rejecting $\mathcal{H}_\textrm{0}$ is called the
\emph{critical region}, while its complement is called the \emph{acceptance
  region}. In the constructed test the critical region is
completely defined by $x_{\textrm{c}}$. If the background expectation $b$
was completely known, we could find $x_{\textrm{c}}$ by solving
\begin{equation}
\label{directp}
P(n\ge x_{c}|b) \equiv \sum_{n=x_{c}}^{\infty} P(n | b)  \le   \alpha, %\leq 1 -\!,
\end{equation}
where $P(n | b)$ is the Poisson distribution, but in general the background
expectation is unknown and so we find the critical value
by inverting the profile likelihood method. Remembering that confidence
regions are constructed such that the true but unknown signal strength $S$ is
outside the confidence region with probability $\alpha$ for any fixed $S$ we
assume the hypothesis $\mathcal{H}_\textrm{0}$ which means $S=0$. The critical
region is therefore defined as the subset of values $x$ which gives rise to
limits not covering $S=0$. That is, $\mathcal{H}_\textrm{0}$ is rejected for
observations that lead to lower limits $s_L$ larger than zero. The limits are
monotonic in $x$, so the hypothesis test is completely characterised by a
critical number $x_\textrm{c}$, and written $x\geq
x_\textrm{c}$. This critical number algorithm is implemented as the method
\textsf{GetCriticalNumber(int\& $n_\textrm{c}$)}.

\subsection{Analysis optimization for signal discovery}

Assuming a specific signal strength $s=S$, it is relevant to consider the probability
of making a discovery. This is given by the \emph{power} of the hypothesis
test, $F_\beta \equiv 1 - \beta$. A signal hypothesis
$\mathcal{H}_{s_\textrm{th}}$ is said to be at the \emph{visibility threshold}
if it leads to a discovery with a pre-specified probability $F_\beta$, for
example 50\%. Discovery is claimed when $x_{{0}}\ge x_\textrm{c}$, so in order
to minimise the visibility threshold, signal is added to the
(background) expectation until the probability for $x\ge x_\textrm{c}$ is at
least $F_\beta$.

\begin{figure}[htb!]
\psfrag{bg}{$P(n|b)$}
\psfrag{bs}{$P(n|b+s_\textrm{th})$}
\psfrag{text3}{$P(n<x_\textrm{c}|b+s_\textrm{th})<\beta$}
\psfrag{text1}{$P(n\ge{}x_\textrm{c}|b)<\alpha$}
\psfrag{xlab}{$x$}
\psfrag{nc}{$x_\textrm{c}$}
\psfrag{ylab}{Probability}
\centering\includegraphics[width=0.8\textwidth]{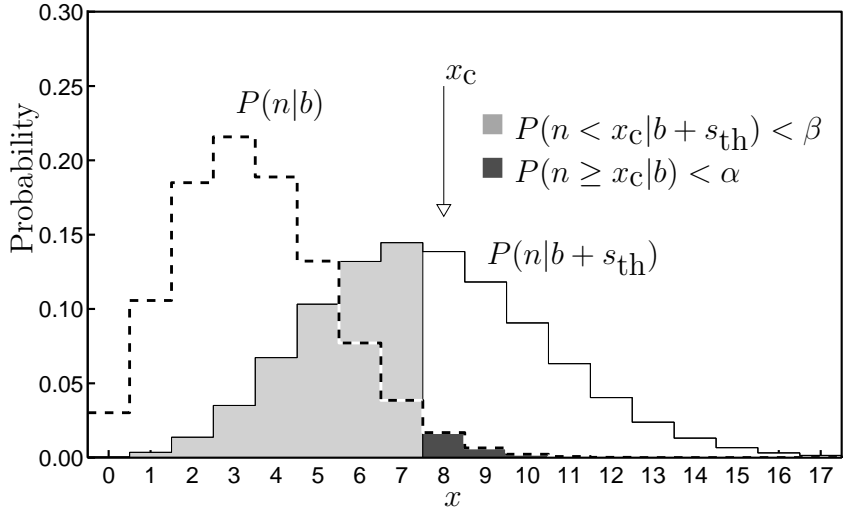}\\
\caption{
For a predefined $\beta$, the visibility
  threshold $s_\textrm{th}$ is the smallest signal that is discovered
  with at least probability $F_\beta =1-\beta$ at significance $\alpha$.
In this example, 
  $\alpha=1\%,\beta=50\%,b=3.5,x_\textrm{c}=8,s_\textrm{th}=4.17$.\label{discpic}}
\end{figure}

For the case of vanishing uncertainties, %in background flux and signal
the visibility threshold can be directly calculated\cite{Hill:2003jk}
from the Poisson distribution by finding the smallest signal 
$s_\textrm{thP}$ fulfilling
\begin{equation}
 P(n \ge x_\textrm{cP}|b+s_\textrm{thP}) \ge F_\beta
\label{simplecrit}
\end{equation}
or equivalently $ P(n< x_\textrm{cP}|b+s_\textrm{thP}) < \beta$, where the critical
value $x_\textrm{cP}$ is that found using equation \ref{directp}. The quantity
$s_\textrm{thP}$ is the visibility threshold for the signal expectation in
case of vanishing uncertainties. The construction is shown in
figure~\ref{discpic}.
\label{optimdisc}
\label{discoverexpl}
Uncertainties are accounted for through the
critical number $x_\textrm{c}(\alpha,b,\Delta_b)$ as function of significance
and expectation number. A method similar to this has previously been described by 
Punzi\cite{punzi}. As in equation \ref{simplecrit}, signal is added to the
(background) expectation until the probability for rejection of
$\mathcal{H}_\textrm{0}$ is at least $F_\beta$. This means

\begin{equation}
\label{critequ}
    x_\textrm{c}\left(F_\beta,b+s, \Delta_{b + s}\right) \ge x_\textrm{c}\left(\alpha,b,\Delta_b\right) 
\end{equation}
where $\Delta_b$ and $\Delta_{b+s}$ represent the total uncertainties of
background, and background plus signal respectively.  Equation \ref{critequ}
is solved numerically by finding the smallest allowed signal expectation $s$
and the solution is called $s_\textrm{th}$. Since the tested hypothesis
$\mathcal{H}_\textrm{0}$ assumes exactly $S=0$, we do not include any
uncertainty in the signal efficiency, while here the background
estimate is assumed Gaussian. The described procedure for finding the critical
number in the presence of uncertainties is thus a function on the form $
x_\textrm{c}(\alpha,b,\Delta_b)$, where $\Delta_b$ is the background
uncertainty.

Assuming that the uncertainties of signal efficiency and the background
estimate are sufficiently uncorrelated and Gaussian (or exact), equation
\ref{critequ} becomes
\begin{equation}
  \label{criteqcrit}
  x_\textrm{c}\left(F_\beta,b(c)+s(c),\sqrt{\Delta_b(c)^2 +
    \Delta_s(c)^2}\right)
\ge   x_\textrm{c}\left(\alpha,b(c),\Delta_b(c)\right). 
\end{equation}

For the observation of $x$ events the \emph{model rejection factor} $\xi(x)$
is defined as
\begin{equation}
  \xi(x) = s_{90}(x_{{0}},b)/s_{\textrm{test}}, 
\end{equation}
where $s_{\textrm{test}}$ is, as in section \ref{optimdisc}, the
expectation number of signal events for an assumed test signal. 
The optimal cut $c$ and the corresponding critical number $x_\textrm{c}$ is
found by minimising the signal strength 
$\mu_{s_{\textrm{th}}}$ as function of the cut $c$. The visibility
threshold for the expected number of observed signal events for a
fixed cut $c$ is
\begin{equation}
s_\textrm{th}=\mu_{s_{\textrm{th}}} \epsilon_s(c).
\label{stheqn}
\end{equation}
The physical threshold signal strength is found in terms of
the test spectrum (in analogy with equation \ref{mrfanalogy}) by 
$ a_\textrm{th}  = a_{\textrm{test}}\cdot {s_\textrm{th}}/{s_\textrm{test}} 
$, or equivalently 
\begin{equation}
 \frac{\mathrm{d}\Phi_\textrm{th}(E)}{dE}  =  \eta \cdot
 \frac{\mathrm{d}\Phi_\textrm{test}(E)}{dE}, 
\label{phystestf}
\end{equation}
where $\eta = s_\textrm{th}/s_{\textrm{test}}$ is the \emph{model detection
  potential}. Minimizing $\eta$ optimizes the analysis such that the signal 
strength required for detection (with at least probability $F_\beta=1-\beta$) is minimized. 
Our code provides the critical number and the
$s_\textrm{th}$ through \textsf{GetCriticalNumber(int\& $n_\textrm{c}$)} and 
%
% This newline was requested by reviewer:
\newline 
\textsf{ bool TRolke2::GetLeastDetectableSignal(\DDDA{} $s_\textrm{th}$, \DDD{}$\beta$)}.

\label{optimsection}
\label{lundbergstat}

\section{Class interface and use}
\label{classdoc}

The library allows seven combinations of efficiency and background rate
models, each presented here. Once the model and its parameters are specified,
the user can obtain limits, critical numbers and so on as explained in 
the subsequent sections.

\subsection{Model Specification methods}

\subsubsection{\textsf{SetGaussBkgGaussEff$(x,bm,em,sde,sdb)$}} 
\begin{tabular}{@{\hspace{0.03\textwidth}}  p{0.97\textwidth}}
      Background: Gaussian,
      Efficiency: Gaussian  

      This model implements the case of Gaussian background with expectation $bm$ and
      standard deviation $sdb$ and Gaussian efficiency with expectation $em$
      and standard deviation $sde$. The integer $x$ is the number of observed
      events.
\end{tabular}

\subsubsection{\textsf{SetGaussBkgKnownEff$(x,bm,sdb,e)$}} 
\label{SetGaussBkgKnownEff}
\begin{tabular}{@{\hspace{0.03\textwidth}}  p{0.97\textwidth}}
      Background: Gaussian, 
      Efficiency: Known     

      This model implements the case of Gaussian background with expectation $bm$ and
      standard deviation $sdb$ and known efficiency $e$. The integer $x$ is the number of observed
      events.
\end{tabular}

\subsubsection{\textsf{SetKnownBkgGaussEff$(x,em,sde,b)$}} 
\label{SetKnownBkgGaussEff}
\begin{tabular}{@{\hspace{0.03\textwidth}}  p{0.97\textwidth}}
      Background: Known, 
      Efficiency: Gaussian   

      This model implements the case of Gaussian efficiency with expectation $em$ and
      standard deviation $sde$ and known background $b$. The integer $x$ is the number of observed
      events. 
\end{tabular}

\subsubsection{\textsf{SetKnownBkgBinomEff$(x,z,b,m)$}} 
\label{SetKnownBkgBinomEff}
\begin{tabular}{@{\hspace{0.03\textwidth}}  p{0.97\textwidth}}
      Background: Known,
      Efficiency: Binomial   

      This model implements the case of known background expectation $b$ and
      Binomial signal efficiency. The integer $z$ is the number of observed
      events (in the signal region) out of the $m$ evaluated signal
      (Monte Carlo) events. The integer $x$ is the number of observed
      events.
%//      m = number of MC events generated
%      z = number of MC events observed
\end{tabular}

\subsubsection{\textsf{SetPoissonBkgKnownEff$(x,y,\tau,e)$}} 
\begin{tabular}{@{\hspace{0.03\textwidth}}  p{0.97\textwidth}}
      Background: Poisson,
      Efficiency: Known    

\label{SetPoissonBkgKnownEff}
The background is either measured simultaneously with
signal, from sidebands, or with separate background Monte Carlo. The real
value $\tau$ is the size of the background region in terms of the size of the
background regions. It can be used in two ways - Either it's 
the ratio between the size of the background and the signal regions
in case background is observed (from sidebands), or in case background is determined from
simulations; the ratio between
simulated and observed exposure time. 
The background in the signal region is estimated from $\tau$ and the integer
$y$, the number of observed events in background region. The integer $x$
is the number of observed events; as always in the signal region.
\end{tabular}

\subsubsection{\textsf{SetPoissonBkgBinomEff$(x,y,z,\tau,m)$}} 
\label{SetPoissonBkgBinomEff}
\begin{tabular}{@{\hspace{0.03\textwidth}}  p{0.97\textwidth}}
      Background: Poisson,
      Efficiency: Binomial

      This model implements the case of Binomial signal efficiency and
      Poissonian background estimate. 
%The integer $z$ is the number of observed
%      events (in the signal region) out of the $m$ evaluated background 
%      (Monte Carlo) events.
      For an explanation of Binomial efficiencies, please refer
      to section \ref{SetKnownBkgBinomEff}, and for Poissonian backgrounds
      to section \ref{SetPoissonBkgKnownEff}.  The integer $x$ is the number of observed
      events.
\end{tabular}

\subsubsection{\textsf{SetPoissonBkgGaussEff$(x,y,em,sde,\tau)$}} 
\begin{tabular}{@{\hspace{0.03\textwidth}}  p{0.97\textwidth}}
      Background: Poisson,
      Efficiency: Gaussian 

      This model implements the case of Gaussian signal efficiency and
      Poissonian background estimate. For an explanation of Binomial
      efficiency, please refer to \ref{SetPoissonBkgBinomEff}, and 
      for Poissonian backgrounds to section \ref{SetKnownBkgBinomEff}.  The integer $x$ is the number of observed
      events.
\end{tabular}

\subsection{Configuration methods and constructor}

The confidence level (CL) is set either at object construction via an optional
argument or with either of the \textsf{SetCL} or \textsf{SetCLSigmas}
methods.

Two options are offered to deal with cases where the maximum likelihood
estimate (MLE) is not in the physical region. 
Bounding is
controled with the \textsf{SetBounding} method. The ``bounded likelihood''
option corresponds to the ``bounds for the physical region'' option in MINUIT/MINOS\cite{minuitA}\cite{minuitB}.
Unbounded likelihood allows the maximum likelihood estimate to be in the
unphysical region. It has better coverage\cite{trolkenull:2004} and is used by default.

\subsection{Limit calculation methods}

The calculation of limits for the model and parameters as specified, 
is performed with the any of the following methods;

\subsubsection{\textsf{   bool GetLimits(\DDDA{} $s_\textrm{L}$, \DDDA{} $s_\textrm{U}$)}}
\begin{tabular}{@{\hspace{0.03\textwidth}}
  p{0.97\textwidth}}
  This method calculates and returns the upper and lower limits for the
  prespecified model, confidence level and model parameters.
\end{tabular}

\subsubsection{\textsf{   bool GetSensitivity(\DDDA{} $s_\textrm{L}$, \DDDA{} $s_\textrm{U}$)}}
\begin{tabular}{@{\hspace{0.03\textwidth}}  p{0.97\textwidth}}
  This method returns the average upper and average lower limits assuming
  vanishing signal. The summation is a Poisson sum over the background
  expectation. This can be used for cut optimization as described in
sectioin \ref{mrf}.
\end{tabular}

\subsubsection{\textsf{bool GetLimitsQuantile(\DDDA{} $s_\textrm{L}$, \DDDA{} $s_\textrm{U}$, int\& $\textrm{out}\_x$, \DDD{}$q = 0.5$)}}
\begin{tabular}{@{\hspace{0.03\textwidth}}  p{0.97\textwidth}}
  This method returns the upper and lower limits for the outcome corresponding to
  a given quantile $q$ assuming vanishing signal and a simple Poisson summation
  using the background expectation. As a default, the quantile value 0.5 is used,
  corresponding to median limits. The quantile and median method has the
  advantage over the sensitivity that it is independent of the signal
  parameter metric. The quantile $x$ value is returned as $\textrm{out}\_x$
\end{tabular}

\subsubsection{\textsf{bool GetLimitsML(\DDDA{} $s_\textrm{L}$, \DDDA{} $s_\textrm{U}$, int\& $\textrm{out}\_x$)}}
\begin{tabular}{@{\hspace{0.03\textwidth}}  p{0.97\textwidth}}
  This method provides the upper and lower limits for the most likely outcome ($\textrm{out}\_x$),
  assuming vanishing signal.
\end{tabular}

\subsection{Hypothesis test methods}

These two methods are used for hypothesis testing as described in section
\ref{lundbergstat}.

\subsubsection{\textsf{   bool GetCriticalNumber(int\& $n_\textrm{c}$)}}
\begin{tabular}{@{\hspace{0.03\textwidth}}  p{0.97\textwidth}}
  Get the smallest number of observed events $x$, corresponding to
  rejection of the null hypothesis. 
\end{tabular}

\subsubsection{\textsf{   bool TRolke2::GetLeastDetectableSignal(\DDDA{} $s_\textrm{th}$, \DDD{}$\beta$)}}.
\begin{tabular}{@{\hspace{0.03\textwidth}}  p{0.97\textwidth}}
  Get the smallest signal strength leading to rejection of the null hypothesis
  with probability $\beta$ as described in section \ref{hyptest}. Currently Gaussian as well as vanishing
  uncertainties are supported.
\end{tabular}

\subsection{Availability and prerequisites}

The latest versions of the code, its documentation and examples are freely
available\cite{codelink}. The class makes use of a number of ROOT
\cite{ROOT} routines for standard mathematical functions, the interactive
interface and bindings which makes it easy to use our methods in Python.
Examples of all functionality of the C++ class are included in our code and
demonstrate its use with Python, as interactive C++, and as a compiled
example program.

\bibliographystyle{natbib}

\end{document}